%% file: GDiffuSE.tex
\documentclass{article}
\usepackage{spconf,amsmath,graphicx,hyperref, caption}
\hypersetup{
    colorlinks=true,
    linkcolor=blue,
    filecolor=magenta,      
    urlcolor=cyan,
    }
\usepackage{tikz}
\usetikzlibrary{arrows.meta, positioning, shapes.geometric, calc}

\usepackage{placeins}

\usepackage{siunitx}
\usepackage{multirow}

\usepackage{caption}
\usepackage{algorithm}
\usepackage{algpseudocode}    
\usepackage{booktabs}
\usepackage{amssymb}

\input{shortcuts.tex}

\usepackage{acro}
\acsetup{first-style=long-short} 
\DeclareAcronym{DNN}{short=DNN,long=deep neural network}
\DeclareAcronym{CNN}{short=CNN, long=convolutional neural network}
\DeclareAcronym{ML}{short=ML, long=maximum likelihood}
\DeclareAcronym{vae}{short=VAE, long=variational autoencoder}
\DeclareAcronym{AR}{short=AR, long=autoregressive}
\DeclareAcronym{DDPM}{short=DDPM, long=denoising diffusion probabilistic model}
\DeclareAcronym{ddim}{short=DDIM, long=denoising diffusion implicit model}
\DeclareAcronym{VP-DDPM}{short=VP-DDPM, long=variance-preserving denoising diffusion probabilistic model}
\DeclareAcronym{cdiffuse}{short=CDiffuSE, long=Conditional Diffusion Probabilistic Model for Speech Enhancement}
\DeclareAcronym{SGMSE}{short=SGMSE, long=score-based generative modeling for speech enhancement}
\DeclareAcronym{nmf}{short=NMF, long=non-negative matrix factorization}
\DeclareAcronym{gan}{short=GAN, long=generative adversarial network, short-plural=GANs, long-plural=generative adversarial networks}
\DeclareAcronym{stft}{short=STFT, long=short-time Fourier transform}
\DeclareAcronym{SE}{short=SE, long=speech enhancement}
\DeclareAcronym{SDE}{short=SDE, long=stochastic differential equation, short-plural=SDEs, long-plural=stochastic differential equations}
\DeclareAcronym{ve}{short=VE, long=variance-exploding}
\DeclareAcronym{VAD}{short=VAD, long=voice activity detector}
\DeclareAcronym{pesq}{short=PESQ, long=Perceptual Evaluation of Speech Quality}
\DeclareAcronym{tf}{short=T-F, long=time-frequency}
\DeclareAcronym{elbo}{short=ELBO, long=evidence lower bound}
\DeclareAcronym{psd}{short=PSD, long=power spectral density}
\DeclareAcronym{SNR}{short=SNR, long=signal-to-noise ratio, short-plural=SNRs, long-plural=signal-to-noise ratios}
\DeclareAcronym{sisdr}{short=SI-SDR, long=scale-invariant signal-to-distortion ratio}
\DeclareAcronym{stoi}{short=STOI, long=Short-Time Objective Intelligibility}
\DeclareAcronym{mos}{short=MOS, long=mean opinion score, short-plural=MOS, long-plural=mean opinion scores}
\DeclareAcronym{p.d.f.}{short=p.d.f., long=probability density function, long-plural=probability density functions}
\DeclareAcronym{GDiffuSE}{short=GDiffuSE, long=guided diffusion for speech enhancement}

\title{GDiffuSE: Diffusion-based Speech Enhancement with Noise Model Guidance}
\twoauthors
 {Efrayim Yanir, David Burshtein}
	{Tel-Aviv University\\
efrayimyanir@mail.tau.ac.il\\
    burstyn@tauex.tau.ac.il
    }
 {Sharon Gannot}
	{Bar-Ilan University\\
Sharon.Gannot@biu.ac.il
	}
\begin{document}
\ninept
\maketitle
\begin{abstract}
We introduce a novel \ac{SE} approach based on a \ac{DDPM}, termed \ac{GDiffuSE}. In contrast to conventional methods that directly map noisy speech to clean speech, our method employs a lightweight helper model to estimate the noise distribution, which is then incorporated into the diffusion denoising process via a guidance mechanism. This design improves robustness by enabling seamless adaptation to unseen noise types and by leveraging large-scale \acp{DDPM} originally trained for speech generation in the context of \ac{SE}. We evaluate our approach on noisy signals obtained by adding noise samples from the BBC sound effects database to LibriSpeech utterances, showing consistent improvements over state-of-the-art baselines under mismatched noise conditions.
\end{abstract}
\section{Introduction}
\label{sec:intro}
Dominant approaches for \ac{SE} utilize discriminative models that map noisy inputs to clean targets~\cite{wang2018supervised}.  These models perform well under matched conditions, but generalize poorly to unseen noise or acoustic environments, often introducing artifacts. 
Generative models that learn an explicit prior over clean speech have gained popularity in recent years, particularly in the context of \ac{SE}.

Diffusion-based generative models~\cite{sohl2015deep,ho2020denoising} gradually add Gaussian noise in a forward process and learn a network to reverse it by iterative denoising. Unlike \acp{vae}, they have no separate encoder---the ``latent'' at step $t$ is the noisy sample itself---and the network learns the score (gradient of log-density) across noise levels~\cite{song2021sde}. They have exhibited promising results in audio generation. For example, DiffWave achieves high-fidelity audio generation with a small number of parameters~\cite{kong2021diffwave}. Recent works adapt diffusion models to \ac{SE}~\cite{welker2022speech,serra2022universal,Lu2021}.
Two main designs have emerged. (i) A \emph{conditioner vocoder} pipeline, where a diffusion vocoder resynthesizes speech utilizing features predicted from the noisy input, with auxiliary losses pushing those features toward clean targets~\cite{serra2022universal,koizumi2022specgrad}.
These methods require an auxiliary loss and use two
separate models for generation and denoising. (ii) \emph{Corruption-aware diffusion} that integrates the corruption model into the forward chain so its reversal directly yields the enhanced signal via linear interpolation between clean and noisy waveforms, e.g.,  CDiffuSE~\cite{Lu2022CDiffuSE}, or by embedding noise statistics in a \ac{SDE} drift~\cite{welker2022speech}. The latter design better reflects real-world, non-white noise~\cite{vincent2011connection}. A recent contribution to the field is the \ac{SGMSE} family of algorithms~\cite{welker2022speech, Richter2023SGMSE, lemercier2023storm_taslp}, which learns a score function that enables sampling from the posterior distribution of clean speech given the noisy observation in the complex \ac{stft} domain. 
All of these methods demonstrate that a conditioned diffusion generator can achieve state-of-the-art performance across diverse noise conditions. However, they all require specialized training of the heavy diffusion model for each type of expected noise.


In this paper, we introduce \ac{GDiffuSE}, a diffusion probabilistic approach to \ac{SE}. \Ac{GDiffuSE} uses the guidance mechanism~\cite{dhariwal2021diffusion} by a lightweight noise model, which guides the signal generated by the DiffWave~\cite{serra2022universal} model towards the estimated clean speech.
Unlike \cite{Iashchenko2023UnDiff}, which relies on reconstruction guidance for known operators, and \cite{nortier2023unsupervisedse}, which uses \ac{nmf}-based linear constraints, \ac{GDiffuSE} introduces test-time training, in which an auxiliary network learns nonlinear noise profiles to guide a frozen diffusion model.
%
That is, given a new unknown noise sample, only the compact noise model has to be trained, which is substantially easier than learning the full distribution of noisy speech. As a result, the system rapidly adapts to unseen acoustic conditions with few noise samples, provided that the noise statistics has not significantly changed between train and inference time.


Our main contributions are threefold: (1) We derive a novel approach for using \ac{DDPM} guidance for \ac{SE} by applying guidance directly into a noise-distribution model for \ac{SE}. (2) We propose a novel reverse process that leverages a foundation diffusion model for \ac{SE}, offering robust adaptability to unseen noise types—assuming the noise statistics remain consistent between the available noise-only utterance and the noise encountered at inference. (3) The experimental results confirm the effectiveness of \ac{GDiffuSE}, achieving improved robustness to mismatched noise conditions compared to related generative \ac{SE} methods.

\section{Problem Formulation}
\label{sec:problem}

Let $y_i = x_{0,i} + w_i$ denote the noisy signal received by a single microphone, 
where $x_{0,i}$ is the clean speech component and $w_i$ is the noise component, 
for $i \in \{0,\ldots,N-1\}$, and $N$ the number of samples in the utterance. Stacking the $N$ samples into column vectors yields
$\mathbf{x}_0 \triangleq (x_{0,i})_{i=0}^{N-1},\quad
 \mathbf{w} \triangleq (w_i)_{i=0}^{N-1},\quad
 \mathbf{y} \triangleq (y_i)_{i=0}^{N-1}$,
leading to the following vector form:
\begin{equation}
  \mathbf{y} = \mathbf{x}_0 + \mathbf{w}.
  \label{eq:sampledaddednoise}
\end{equation}
Given $\mathbf{y}$, the goal of the \ac{SE} algorithm is to estimate
$\hat{\mathbf{x}} \triangleq (\hat{x}_i)_{i=0}^{N-1}$
that is perceptually and/or objectively close to $\mathbf{x}_0$.

\section{Proposed Method}
\label{sec:format}
In this section, we derive the proposed \ac{SE} algorithm. Sec.~\ref{sect:conditional_sampling_for_DDPM} presents the use of DDPM guidance for \ac{SE}, and Sec.~\ref{sect:noisemodel} describes the training of the noise model that guides the DDPM. The complete process is illustrated in Fig.~\ref{fig:NGDiffuSE}.

\subsection{DDPM Guidance for Speech Enhancement} 
\label{sect:conditional_sampling_for_DDPM}
\ac{DDPM}~\cite{ho2020denoising} uses a diffusion processs~\cite{sohl2015deep} for generative sampling. DDPM guidance~\cite{dhariwal2021diffusion} modifies the standard generative sampling procedure of \ac{DDPM} to a conditional one as summarized in \cite[Algorithm 1]{dhariwal2021diffusion}. We suggest adopting this approach for \ac{SE} in a new way, using guidance from the noise model distribution, as summarized in Algorithm~\ref{alg:Inference}.

We follow the notations in  \cite{ho2020denoising,dhariwal2021diffusion}. The data distribution of the clean speech is given by $\bx_0 \sim q(\bx_0)$. In the forward diffusion process, a Markov chain progressively adds noise to $\bx_0$ to produce $\bx_1,\bx_2,\ldots,\bx_T$ as follows:
\begin{equation}
\bx_t = \sqrt{1-\beta_t} \bx_{t-1} + \beta_t \be_t,\, \be_t \sim \mathcal{N}(\mathbf{0}, \mathbf{I}),\be_t \perp\!\!\!\perp \bx_{t-1},
\label{eq:diffevolution}
\end{equation}
where $\be_t$ (Gaussian distributed with zero mean, and identity covariance matrix) is statistically independent of $\bx_{t-1}$, and $\beta_t \in [\beta_{\text{start}}, \beta_{\text{end}}]$ is a schedule parameter. Other schedule parameters, $\alpha_t$ and $\overline{\alpha}_t$ are defined in \cite{ho2020denoising,dhariwal2021diffusion} in the following way:
\begin{equation}
\alpha_t = 1 - \beta_t ,\quad \text{ } \bar{\alpha}_t = \prod_{s=1}^{t} \alpha_s = \prod_{s=1}^{t} (1 - \beta_s).
\label{eq:alpha}
\end{equation}
Consequently, the $t$-step marginal is \cite{ho2020denoising}:
\begin{equation}
  \bx_t \;=\; \sqrt{\bar{\alpha}_t}\,\bx_0 + \sqrt{1-\bar{\alpha}_t}\,\hat{\be}_t,
  \; \hat{\be}_t \sim \mathcal{N}(0,\mathbf{I}),\hat{\be}_t \perp\!\!\!\perp \bx_{0}.
  \label{eq:xt_marginal}
\end{equation}
Denoising is performed by recursively applying the following reverse process, for $t=T,T-1,\ldots,1$:
\begin{equation}\label{eq:denoisin_dist}
  p_{\boldsymbol{\theta}}(\bx_{t-1}\!\mid\!\bx_t)
  = \mathcal{N}\!\big(\bx_{t-1};\,\bmu(\bx_t,t),\,\sigma_t^2 \mathbf{I}\big).
\end{equation}
Since the distribution of the reverse process is intractable, it is modeled by a \ac{DNN}, where $\btheta$ represents the set of trainable parameters of the denoising network. Therefore, sampling can be expressed with:
\begin{equation}
  \bx_{t-1} \;=\; \bmu(\bx_t,t) + \sigma_t\,\bz_t, \; \bz_t \sim \mathcal{N}(\mathbf{0}, \mathbf{I}),\;\; \bz_t \perp\!\!\!\perp \bx_t,
\label{eq:denoising}
\end{equation}
where the mean $\bmu(\bx_t,t)$ can be expressed using the standard noise-prediction form:
\begin{equation}
\bmu(\bx_t,t) = \frac{1}{\sqrt{\alpha_t}} 
\left( \bx_t - \frac{1-\alpha_t}{\sqrt{1-\overline{\alpha}_t}} \bepsilon_{\btheta}(\bx_t,t) \right).
\label{eq:mu_diffwave}
\end{equation}
The function $\bepsilon_{\btheta}(\bx_t,t)$ is the network’s estimate of the injected noise~\cite[Algorithm 1]{ho2020denoising}.
As shown in \cite{ho2020denoising} and \cite{kong2021diffwave}, for accelerating the computation it is useful to use in \eqref{eq:denoising}:
\begin{align}
\sigma_t^2 \;=\tilde{\beta}_t = 
\begin{cases}
\frac{1 - \bar{\alpha}_{t-1}}{1 - \bar{\alpha}_t} \, \beta_t & \text{for } t > 1 \\
\beta_1 & \text{for } t = 1
\end{cases}
\,.
\end{align}
For an \ac{SE} problem, we want to add a guidance component to guide the diffusion process towards the clean speech $\by$. For that, we train the diffusion model in the standard way, but then we wish to sample $\bx_0$ from the conditional \ac{p.d.f.} $p_{\boldsymbol{\phi}}(\bx_0 \given \by)$, modeled by a \ac{DNN} $\boldsymbol{\phi}$. This can be done as described in \cite{sohl2015deep,dhariwal2021diffusion}. Rather than using \eqref{eq:denoising}-\eqref{eq:mu_diffwave} we use:
\begin{equation}
\bx_{t-1} =\boldsymbol{\mu}_t^{\mathrm{guid}} + \sigma_t \tilde{\be}_t, \qquad \tilde{\be}_t \sim \mathcal{N}(\mathbf{0}, \mathbf{I}).
\label{eq:denoising_cond}
\end{equation}
where
\begin{equation}
\boldsymbol{\mu}_t^{\mathrm{guid}} = \bmu(\bx_t,t) + s_t \frac{\beta_t}{\sqrt{\alpha_t}} \nabla_{{\bx}} \log p_{\boldsymbol{\phi}}(\by\given{\bx}) |_{\\{\bx}=\bmu({\bx}_t,t)}
\: .
\label{eq:mu_cond}
\end{equation}
We set the gradient scale, $s_t$, according to the schedule: 
\begin{equation}
s_t
\;=\;
\lambda_{\max}\!\left(\frac{\sqrt{1-\bar\alpha_t}}{\sqrt{1-\bar\alpha_{1}}}\right)^{\gamma},\qquad \gamma>0,\; \lambda_{\max}>0.
\label{eq:guid_parameters}
\end{equation}
Intuitively, this schedule applies stronger guidance at early, low-\ac{SNR}
steps—when the estimate is dominated by noise—and gradually relaxes the guidance as the
effective \ac{SNR} increases, allowing the generative model to refine fine details without
over-constraining the reconstruction.
This schedule is consistent with standard \ac{SNR}-dependent scheduling for diffusion models~\cite{Karras2022EDM}, and aligns with recent evidence that guidance strength should vary with the noise level rather than remain constant~\cite{Wang2024CFG,Kynkaanniemi2024Interval}.

Now, given observation $\by$ we can use~\eqref{eq:denoising_cond}-\eqref{eq:mu_cond} to estimate the clean speech. We just need to know $\nabla_{\bx_t} \log p_{\boldsymbol{\phi}}(\by\given\bx_t)$.

\usetikzlibrary{fit} 

\begin{figure}[ht]
\centering
\begin{tikzpicture}[
  scale=0.7, transform shape,      
  font=\small,
  box/.style = {draw, rounded corners=2pt, minimum width=30mm,
                minimum height=6.5mm, align=center},
  smallbox/.style = {draw, rounded corners=2pt, minimum width=24mm,
                     minimum height=5.5mm, align=center, font=\scriptsize},
  bigsmallbox/.style = {draw, rounded corners=2pt, minimum width=28mm,
                        minimum height=7mm, align=center, font=\footnotesize},
  arrow/.style = {->, >=Latex, thick},
  stage/.style = {draw, thick, inner sep=4pt, rounded corners=4pt}
]

\node[box, fill=blue!10] (noise) at (0,0)
      {Noise Sample $\bar{\bw}$};

\node[box] (phi_train) at (0,-20mm)
      {Trained Noise Model\\$\phi_t$};

\draw[arrow] (noise) -- (phi_train);

\node[stage, fit=(noise) (phi_train),
      label={[font=\bfseries]north:Training Stage}]
      {};

\node[box, fill=gray!10] (latent) at (80mm,0)
      {$\bx_t$\\Diffusion Model $\theta$};

\node[box] (diff)   at (80mm,-20mm)
      {Diffusion Process};

\node[box, fill=green!10] (clean) at (80mm,-31mm)
      {Cleaner Speech\\$\bx_{t-1}(\boldsymbol{\mu}_t^{\mathrm{guid}},\,\boldsymbol{\Sigma}_t)$};

\node[box, fill=yellow!10] (phi_frozen) at (40mm,-20mm)
      {Frozen Noise Model\\$\phi_t$};

\node[bigsmallbox, fill=orange!10] (noise_est)
      at (40mm,6mm)
      {Noise Estimation\\$\bv_t \gets \by - \dfrac{1}{\sqrt{\overline{\alpha}_t}}\;\boldsymbol{\mu_\theta}(\bx_t,t)$};

\node[bigsmallbox, fill=red!10] (noisy_y)
      at (40mm,18mm)
      {Noisy Speech $\by$};

\draw[arrow] (latent) -- (diff);
\draw[arrow] (phi_frozen) -- (diff);
\draw[arrow] (diff)   -- (clean);

\draw[arrow] (diff.north west) .. controls +(-10mm,8mm) and +(0,-4mm)
      .. node[pos=0.55, above, fill=white, inner sep=1pt] {$\boldsymbol{\mu_\theta}$}
      (noise_est.south east);
\draw[arrow] (noisy_y.south) -- (noise_est.north);
\draw[arrow] (noise_est) -- (phi_frozen);

\draw[arrow]
  (phi_train.east) .. controls +(right:15mm) and +(left:15mm) ..
  (phi_frozen.west);

\node[stage, fit=(latent) (diff) (clean) (phi_frozen) (noise_est) (noisy_y),
      label={[font=\bfseries]north:Inference Stage}]
      {};

\end{tikzpicture}
    \addtolength{\belowcaptionskip}{-16pt}
\caption{%
\textbf{GDiffuSE:} The trained noise model guides the diffusion model for \ac{SE}.  
\textbf{Training stage:} Noise sample $\bar{\bw} \in \mathbb{R}^N$ trains the noise models $\boldsymbol{\phi}_t$ for each $t$.  
\textbf{Inference stage:} Starting from $\bx_t$ (white noise for $t=T$), the diffusion process, guided by the loss from $\phi_t$~\eqref{eq:lossgdiff}, generates ${x}_{t-1}$; the clean estimate is ${x}_0$. The input to $\phi_t$ is the noise estimate (which uses $\by$). This is repeated $T$ times (See Algorithms~\ref{alg:Training},~\ref{alg:Inference}).}
\label{fig:NGDiffuSE}
\end{figure}
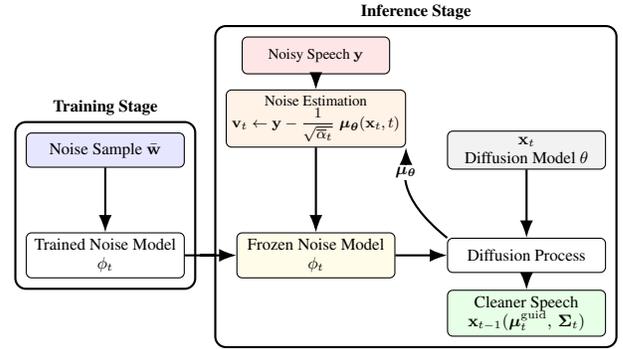

\subsection{Noise Model Training}
\label{sect:noisemodel}
In this section, we specify how to train the noise model, $\boldsymbol{\phi}$.
%
The conditional density $p_{\boldsymbol{\phi}}(\by\given\bx_t)$ is inferred using the noise at the $t$-th guided diffusion step and the additive (acoustic) noise, as follows. 
Combining~\eqref{eq:xt_marginal} with \eqref{eq:sampledaddednoise} yields,
\begin{equation}\label{eq:y_expand}
\by = \bx_0 + \bw = \frac{1}{\sqrt{\bar{\alpha}_t}}\,\bx_t
      - \sqrt{\frac{1-\bar{\alpha}_t}{\bar{\alpha}_t}}\,\hat{\be}_t + \bw .
\end{equation}
Denote the combined noise:
\begin{equation}
\bv_t \defined -\frac{\sqrt{1-\overline{\alpha}_t}}{\sqrt{\overline{\alpha}_t}} \hat{\be}_t + \bw=\bw-g(t)\hat{\be}_t
\label{eq:nt_def}
\end{equation}
where
\begin{equation}
g(t)=\sqrt{\frac{1-\overline{\alpha}_t}{\overline{\alpha}_t}}.
\label{eq:g_def}
\end{equation}
The first component is the diffusion noise, and the second is the acoustic noise that should be suppressed.
Consequently, the conditional probability of the measurements given the desired speech estimate at the $t$-th step is given by
\begin{equation}
p_{\boldsymbol{\phi}}(\by \given \bx_t) = p^{\bV_t|\bX_t}_{\boldsymbol{\phi}}(\by - \frac{1}{\sqrt{\overline{\alpha}_t}} \bx_t \given \bx_t).
\end{equation}
Hence, the required conditional probability density simplifies to the conditional density of the random variable $\bV_t$ given the variable $\bX_t$, $p^{\bV_t|\bX_t}_{\boldsymbol{\phi}}(\bv_t | \bx_t)$.
Obviously, the additive noise, $\bw$, is statistically independent of $\bx_t$.
To further simplify the derivation, we  also make the assumption that $\hat{\be}_t$ is independent of $\bx_t$. Consequently, the density of $\bV_t$ given $\bx_t$ becomes the density of $\bw-g(t)\cdot\hat{\be}_t$, where
 $\hat{\be}_t \sim \cN(\mathbf{0},\mathbf{I})$ is independent of $\bw$.
We also assume the availability of a noise sample $\bar{\bw}$ from the same distribution as $\bw$, which can be used to train a model for $\bv_t$. 
In practice, a \ac{VAD} can be used to allocate such segments from the given noisy utterance.
Given a segment $\bar{\bw}$, for each diffusion step $t$ we can compute $g(t)$, the noise level for a specific step~\eqref{eq:g_def}, and generate noise $\bv_t$ with the required density:
\begin{equation}
v_{t,i} = {\bar{w}}_i - \hat{e}_t \cdot g(t) ,\; \hat{e}_t \sim \cN\left( 0, 1 \right).
\end{equation}
\sloppy For inferring $p^{\bV_t}_{\boldsymbol{\phi}}(\bv_t)$ we apply \ac{ML}. 
The log likelihood is given by:
\begin{equation}
\log P(v_0, \dots, v_{N-1} \mid \theta) = \sum_{i=0}^{N-1} \log p(v_i \mid v_0,\ldots,v_{i-1},\theta),
\end{equation}
and therefore, we need the conditional distribution of $ v_{t,i} \given (v_{t,0},... v_{t,i-1})$. 
We model it by a Gaussian: 
$
v_{t,i} \given (v_{t,0},... v_{t,i-1}) \sim \cN(\cdot,\mu_{t,i}, \sigma^2_{t,i} ).$
The noise is modeled separately for each $t$ with shifted causal \acp{CNN}~\cite{van2016wavenet}
to predict the mean and the variance:
\begin{equation}
\mu_{t,i}(v_{t,0},... v_{t,i-1}), \sigma_{t,i}^2(v_{t,0},... v_{t,i-1}) = \boldsymbol{\phi}_t(v_{t,0},\ldots,v_{t,i-1})
\label{eq:musigeq}
\end{equation}
and $\boldsymbol{\phi}_t$ is trained using the \ac{ML} loss $(-\log L)_t$:
\begin{equation}
\text{loss}_t(\bv_t) = \sum_{i=0}^{N-1} \left[ \log\left( \sqrt{2\pi} \cdot \sigma_{t,i} \right) + \frac{(v_{t,i} - \mu_{t,i})^2}{2 \cdot \sigma_{t,i}^2} \right].
\label{eq:lossgdiff}
\end{equation}
The training of the noise model given a noise sample $\bar{\bw}$ is summarized in Algorithm~\ref{alg:Training}. The guided reverse diffusion is summarized in Algorithm~\ref{alg:Inference}. The training and inference procedures are schematically depicted in Fig.~\ref{fig:NGDiffuSE}.
%
%
%
%
\begin{algorithm}[t]
\caption{Noise Model Training}\label{alg:Training}
\begin{algorithmic}[1]
\Require noise sample $\bar{\bw} \in \mathbb{R}^N$, diffusion steps $T$, \# epochs $E$, step size $\eta$, schedule $g(t)$.
\For{$t \gets T$ \textbf{down to} $1$}
  \State Compute $g(t),\hat{\mathbf{e}}_t \sim \cN(0, {\mathbf{I}})$
  \State $\bv_t \gets \bar{\bw} - \hat{\be}_t\,g(t)$ \Comment{elementwise: $v_{t,i}=\bar{w}_i-\hat{e}_t\,g(t)$}
  \For{$k \gets 1$ \textbf{to} $E$} \Comment{NumEpochs}
    \State $(\mu_{t,i},\sigma_{t,i}^2)_{i=0}^{N-1} \gets \boldsymbol{\phi}_t\!\big(v_{t,0},\ldots,v_{t,i-1}\big)$
    \State $\text{loss}_t(\bv_t) \gets$ See~\eqref{eq:lossgdiff}
    \State $\boldsymbol{\phi}_t \gets \textsc{AdamStep}\, \!\big(\boldsymbol{\phi}_t,\nabla_{\boldsymbol{\phi}_t} \,\text{loss}_t,\eta\big)$
  \EndFor
\EndFor
\State \Return $\{\boldsymbol{\phi}_t\}_{t=1}^{T}$
\end{algorithmic}
\end{algorithm}
\begin{algorithm}[t]
\caption{Guided reverse diffusion (sampling)}\label{alg:Inference}
\begin{algorithmic}[1]
\Require schedules $\{\alpha_t,\bar{\alpha}_t,\tilde{\beta}_t\}$; denoiser $\epsilon_\theta$; noise models $\{\boldsymbol{\phi}_t\}$; scheduled scales $\{s_t\}$; observation $\by$
\State $\bx_T \sim \mathcal{N}(\mathbf{0},\mathbf{I})$
\For{$t \gets T$ \textbf{down to} $1$}
  \State $\boldsymbol{\mu_\theta}(\bx_t,t) \gets \dfrac{1}{\sqrt{\alpha_t}}
        \Big(\bx_t - \dfrac{\beta_t}{\sqrt{1-\bar{\alpha}_t}}\;\epsilon_\theta(\bx_t,t)\Big)$
  \State $\boldsymbol{\sigma_\theta^2}(\bx_t,t) \gets \tilde{\beta}_t$
  \State $\bv_t \gets \by - \dfrac{1}{\sqrt{\overline{\alpha}_t}}\;\boldsymbol{\mu_\theta}(\bx_t,t)$ 
  \State $
  \{\mu_{t,i},\sigma_{t,i}^2\}_{i=0}^{N-1} \gets \boldsymbol{\phi}_t\!\big(v_{t,0},\ldots,v_{t,i-1}\big)$
  \State $\text{loss}_t(\bv_t) \gets$ See~\eqref{eq:lossgdiff}
    \State $\boldsymbol{\mu}_t^{\mathrm{guid}} \gets
       \boldsymbol{\mu_\theta}(\bx_t,t) + s_t\Big(\dfrac{\beta_t}{\sqrt{\alpha_t}}\Big)\Big( -\dfrac{1}{\sqrt{\overline{\alpha}_t}}\dfrac{\partial\,\text{loss}_t(\bv_t)}{\partial \bv_t}\Big)$
    \State $\boldsymbol{\Sigma}_t \gets {\rm diag}\!\big(\boldsymbol{\sigma_\theta^2}(\bx_t,t)\big)$
    \State $\bx_{t-1} \sim \mathcal{N}\!\big(\boldsymbol{\mu}_t^{\mathrm{guid}},\,\boldsymbol{\Sigma}_t\big)$
\EndFor
\State \Return $\bx_0$
\end{algorithmic}
\end{algorithm}
It is important to note that the backbone diffusion model is trained solely on clean speech, so large amounts of noisy data are not required. In practice, we employ a \emph{pretrained} diffusion model for clean speech (see Sec.~\ref{ssec:impl}), and only the lightweight noise model needs to be trained in the proposed scheme.

%
%
%
%
\section{Experimental Study}
\label{sec:results}
In this section, we provide the implementation details of the proposed method, describe the competing method, the datasets used for training and testing, and evaluate the method's performance.

\subsection{Implementation details}
\label{ssec:impl}
The noise model architecture is a \ac{CNN} with 4 causal convolutional layers and linear heads for $\mu_{t,i}$ and $ \sigma_{t,i}$, featuring residual connections and weight normalization. We use a WaveNet-style tanh–sigmoid gate, $\mathrm{Gate}(h,g)=\tanh(h)\odot\operatorname{sigm}(g)$,
with $h=\mathrm{Conv}_{\text{causal}}(x)$ and $g=\mathrm{Conv}_{1\times1}(h)$.
The network's parameters are a kernel size of 9, 2 channels, and dilations of [1, 2, 4, 8]. The parameters $\lambda_{\max}$ and $\gamma$ in~\eqref{eq:guid_parameters} exhibit a wide range of values with good results, spanning between $[0.5, 1]$ for both. We calibrated them on one clip per \ac{SNR} level to $\gamma=0.7$ and \(\lambda_{\max}=[0.8,0.72,0.6,0.55]\) for \ac{SNR} levels [10,5,0,-5]~dB, respectively. Since the noise model is extremely lightweight (172 parameters), the computational overhead is minimal. On an NVIDIA GeForce GTX TITAN X, adaptation incurs 3.10\% overhead (with four GPUs), whereas inference incurs only 0.6\% overhead relative to DiffWave.
For the generator, we used the unconditional DDPM model, trained by UnDiff \cite{Iashchenko2023UnDiff}  with 200 diffusion steps, on the Datasets VCTK~\cite{vctk_0_92_2019} and LJ-Speech~\cite{ljspeech17}.

\subsection{Baseline method} We used SGMSE \cite{Richter2023SGMSE}, a fully generative speech denoising model, as our baseline. This model was trained on clean speech from either the WSJ0 Dataset~\cite{LDC93S6A} or the TIMIT dataset~\cite{LDC93S1}, and on noise signals from the CHiME3 Dataset~\cite{Barker2015CHiME3}. We also tested CDiffuSE~\cite{Lu2022CDiffuSE} as a simpler generative method.

\subsection{Datasets} As the backbone diffusion model is pretrained (with clean speech), we only need noise clips for training the noise model and noisy signals (clean speech plus noise) for inference.
We used LibriSpeech~\cite{panayotov2015librispeech} (out-of-domain) as the clean speech dataset. 
%
For the noise, we selected real clips from the BBC sound effects dataset~\cite{bbc_sound_effects}. This lesser-known corpus was chosen because it includes noise types that are rarely found in widely used datasets such as CHIME3, thereby enabling a more rigorous evaluation of robustness.
%

For the test set, we selected 20 speakers, each contributing a single 5-second clean utterance resampled to 16 kHz. The noise data comprised 25-second recordings, of which 20 seconds were used to train the noise model and the remaining 5 seconds for testing. Noisy utterances were generated by mixing the 5-second clean speech with noise at various \ac{SNR} levels. The noise model was trained using a fixed number of epochs, decreasing from 70 (for $t=0$) to 10 (for $t=T$). A preliminary study suggests that substantially shorter training segments are sufficient.

\subsection{Evaluation metrics} To assess the performance of the proposed \ac{GDiffuSE} algorithm and compare it with the baseline method we used the following metrics: STOI~\cite{Taal2011STOI}, PESQ~\cite{Rix2001PESQ}, SI-SDR~\cite{LeRoux2019ICASSP} (all intrusive metrics that require a clean reference), and DNSMOS~\cite{Reddy2021DNSMOS} (a non-intrusive, reference-free measure).

\subsection{Experimental results}

Results for real noise signals from the BBC sound effects dataset are shown in Table~\ref{tab:resultsbbc_with_input}. Our method consistently outperforms \ac{SGMSE} in PESQ and SI-SDR across all SNR levels, even if the gains are modest. Although \ac{SGMSE} achieves higher STOI and DNSMOS scores, informal listening tests confirm that our approach delivers perceptual sound quality that is noticeably better.

To further assess robustness, we selected 20 additional noise clips with spectral profiles that emphasize higher frequencies. Since the noise statistics remain relatively stable over time, these clips align well with our model assumptions.  
As shown in Table~\ref{tab:results_filteredhigh}, the performance gains of \ac{GDiffuSE} over \ac{SGMSE} become even more pronounced in this setting. More research is required to comprehensively characterize the noise types for which \ac{GDiffuSE} achieves the most significant gains.

\begin{figure}[h]
    \centering
    \includegraphics[width=0.5\textwidth]{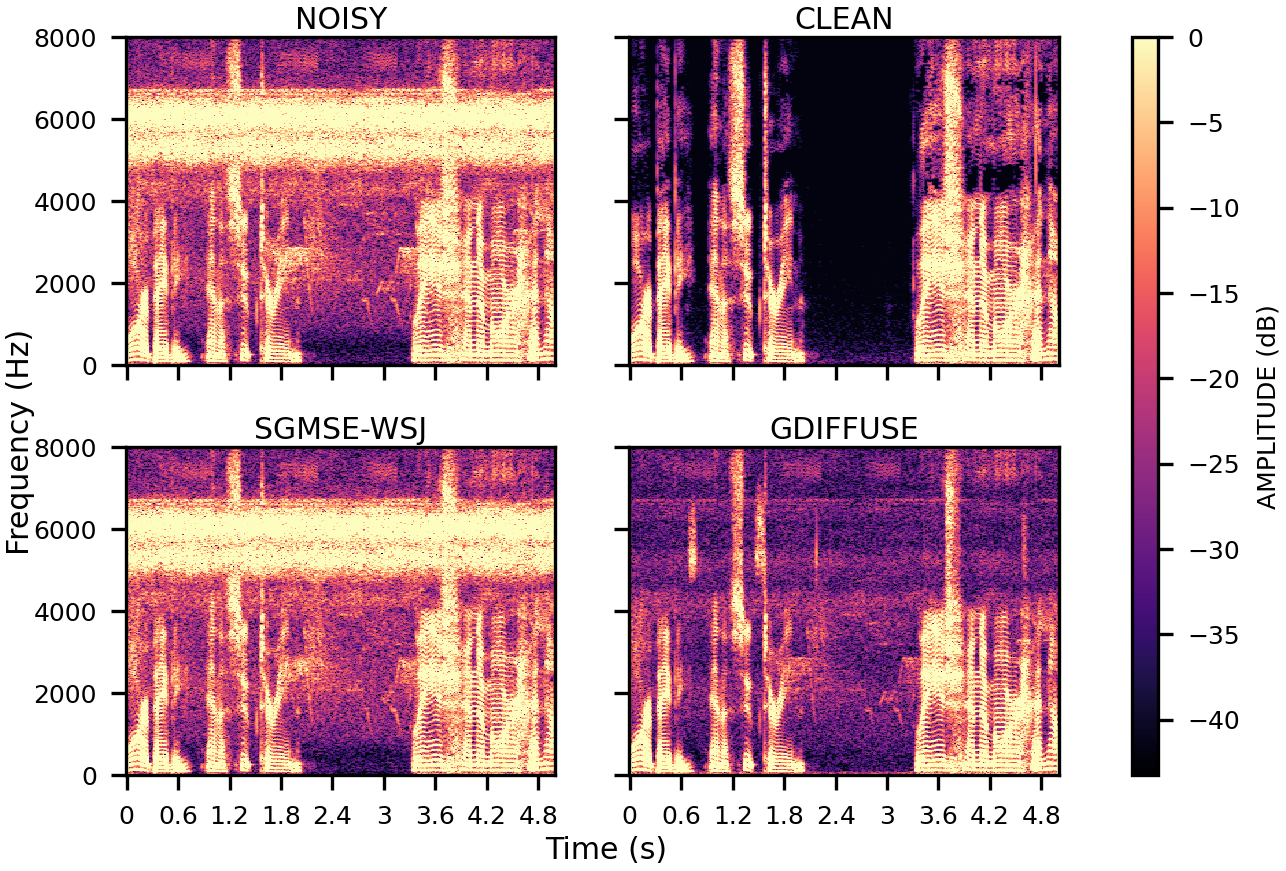}
    \addtolength{\belowcaptionskip}{-16pt}
    \addtolength{\abovecaptionskip}{-18pt}
    \caption{Spectrograms assessment for sample NHU05093027 (monsoon forest) drawn from the BBC sound effect dataset.}
    \label{fig:spectogramcomp_na}
\end{figure}

The comparison of spectrograms in Fig.~\ref{fig:spectogramcomp_na} highlights this difference: while \ac{SGMSE} struggles to suppress the unseen noise, \ac{GDiffuSE} adapts effectively to these challenging conditions.
Audio examples\footnote{\url{https://ephiephi.github.io/GDiffuSE-examples.github.io}}
 further confirm the superiority of the proposed method, particularly for unfamiliar noise types, where improvements in PESQ and SI-SDR are most evident.

%
%
%
%
%

%
%
%
%
%
%
%
%
%
%
%
%

\begin{table}[h]
\caption{Objective evaluation using noise drawn from BBC sound effect dataset (higher is better). `W' stands for WSJ0 and `T' for TIMIT.}
\label{tab:resultsbbc_with_input}
\centering
\scriptsize
\setlength{\tabcolsep}{3pt}
\renewcommand{\arraystretch}{1.05}
\resizebox{\columnwidth}{!}{%
\begin{tabular}{
  c
  l
  S[table-format=1.2(2)]
  S[table-format=1.2(2)]
  S[table-format=1.2(2)]
  S[table-format=2.2(2)]
}
\toprule
\textbf{SNR} & \textbf{Method} & {\textbf{STOI}} & {\textbf{PESQ}} & {\textbf{DNSMOS}} & {\textbf{SI-SDR}} \\
\midrule
\multirow{4}{*}{10}
 & \ac{GDiffuSE}        & 0.91 +- 0.05 & \bfseries 1.60 +- 0.36 & 2.92 +- 0.24 & \bfseries 14.80 +- 3.55 \\
 & sgmseW   & \bfseries 0.94 +- 0.04 & 1.59 +- 0.34 & \bfseries 3.06 +- 0.27 & 14.23 +- 3.07 \\
 & sgmseT  & 0.93 +- 0.04 & 1.46 +- 0.27 & 3.04 +- 0.25 & 12.41 +- 1.77 \\
 & Input       & 0.90 +- 0.06 & 1.20 +- 0.14 & 2.42 +- 0.41 & 10.00 +- 0.02 \\
\midrule
\multirow{4}{*}{5}
 & \ac{GDiffuSE}        & 0.86 +- 0.08 & \bfseries 1.40 +- 0.32 & 2.73 +- 0.32 & \bfseries 10.91 +- 4.47 \\
 & sgmseW   & \bfseries 0.90 +- 0.06 & 1.34 +- 0.30 & \bfseries 2.94 +- 0.27 & 10.46 +- 4.03 \\
 & sgmseT  & 0.88 +- 0.07 & 1.20 +- 0.16 & 2.78 +- 0.27 & 7.80 +- 2.65 \\
 & Input       & 0.84 +- 0.09 & 1.11 +- 0.09 & 2.03 +- 0.46 & 5.01 +- 0.03 \\
\midrule
\multirow{4}{*}{0}
 & \ac{GDiffuSE}        & 0.78 +- 0.11 & \bfseries 1.25 +- 0.27 & 2.65 +- 0.33 & \bfseries 6.66 +- 5.52 \\
 & sgmseW   & \bfseries 0.84 +- 0.10 & 1.18 +- 0.17 & \bfseries 2.79 +- 0.34 & 6.04 +- 4.68 \\
 & sgmseT  & 0.82 +- 0.10 & 1.11 +- 0.09 & 2.61 +- 0.31 & 3.38 +- 3.53 \\
 & Input       & 0.77 +- 0.11 & 1.07 +- 0.06 & 2.41 +- 1.05 & 0.02 +- 0.04 \\
\midrule
\multirow{4}{*}{-5}
 & \ac{GDiffuSE}        & 0.69 +- 0.15 & \bfseries 1.12 +- 0.15 & 2.26 +- 0.61 & \bfseries 1.34 +- 6.42 \\
 & sgmseW   & \bfseries 0.76 +- 0.14 & 1.09 +- 0.10 & \bfseries 2.51 +- 0.39 & 0.77 +- 5.52 \\
 & sgmseT  & 0.74 +- 0.14 & 1.07 +- 0.06 & 2.35 +- 0.36 & -1.46 +- 4.24 \\
 & Input       & 0.69 +- 0.13 & 1.09 +- 0.17 & 2.04 +- 1.03 & -4.97 +- 0.07 \\
\bottomrule
\end{tabular}%
}
\end{table}

 \begin{table}[!h]
\caption{Evaluation on 20 samples with spectral profile emphasizing high frequencies at SNR=5\,dB.}
\label{tab:results_filteredhigh}
\centering
\scriptsize
\setlength{\tabcolsep}{3pt}
\renewcommand{\arraystretch}{1.05}
\resizebox{\columnwidth}{!}{%
\begin{tabular}{
  l
  S[table-format=1.2(2)]
  S[table-format=1.2(2)]
  S[table-format=1.2(2)]
  S[table-format=2.2(2)]
}
\toprule
\textbf{Method} & {\textbf{STOI}} & {\textbf{PESQ}} & {\textbf{DNSMOS}} & {\textbf{SI-SDR}} \\
\midrule
GDiffuSE         & 0.88 +- 0.07 & \bfseries 1.39 +- 0.24 & \bfseries 2.87 +- 0.25 & \bfseries 11.25 +- 3.21 \\
sgmseWSJ0    & \bfseries 0.91 +- 0.07 & 1.26 +- 0.17 & 2.82 +- 0.25 & 9.43 +- 2.64 \\
sgmseTIMIT   & 0.89 +- 0.07 & 1.20 +- 0.14 & 2.84 +- 0.29 & 8.64 +- 2.85 \\
CDiffuSE     & 0.80 +- 0.06 & 1.12 +- 0.07 & 2.31 +- 0.46 & 3.66 +- 3.23 \\
\midrule
Input        & 0.85 +- 0.09 & 1.07 +- 0.03 & 1.98 +- 0.47 & 5.00 +- 0.03 \\
\bottomrule
\end{tabular}%
}
\end{table}

\section{conclusions}
In this work, we introduce \ac{GDiffuSE}, a lightweight \ac{SE} method that leverages a pre-trained diffusion foundation model as its backbone and circumvents retraining it by employing a small guidance mechanism. 
By modeling the noise distribution---an easier task than mapping noisy to clean speech---our approach requires only a short reference noise clip, assuming stable noise statistics between training and inference, thereby improving robustness to unfamiliar noise types. For noise signal types that were not encountered during \ac{SGMSE} training, our method surpasses the state-of-the-art \ac{SGMSE}, as demonstrated by our experimental study and the project website.

\bibliographystyle{IEEEtran}
\bibliography{venues_longer,refs}

\end{document}

%% file: shortcuts.tex
\newcommand{\rem}[1]{}

\newcommand{\bre}{\begin{equation}}
\newcommand{\ere}{\end{equation}}
\newcommand{\be}{{\bf {e}}}
\newcommand{\bra}{\begin{eqnarray}}
\newcommand{\era}{\end{eqnarray}}
\newcommand{\bfg}{\begin{figure}[hbtp]}
\newcommand{\efg}{\end{figure}}
\newcommand{\bver}{\begin{verbatim}}
\newcommand{\ever}{\end{verbatim}}
\newcommand{\bit}{\begin{itemize}}
\newcommand{\eit}{\end{itemize}}
\newcommand{\ben}{\begin{enumerate}}
\newcommand{\een}{\end{enumerate}}

\newcommand{\bepsilon}{\boldsymbol\epsilon}

\newcommand{\btheta}{\boldsymbol\theta}

\newcommand{\bmu}{\boldsymbol\mu}

\newcommand{\given}{\: | \:}

\newcommand{\baa}{\begin{eqnarray*}}
\newcommand{\eaa}{\end{eqnarray*}}

\newcommand{\bX}{{\bf X}}

\newcommand{\bV}{{\bf V}}

\newcommand{\bv}{{\bf v}}

\newcommand{\bw}{{\bf w}}

\newcommand{\bx}{{\bf x}}
\newcommand{\by}{{\bf y}}
\newcommand{\bz}{{\bf z}}

\newcommand{\cN}{{\mathcal{N}}}

\newcommand{\defined}{\triangleq}

\def\defined{\: {\stackrel{\scriptscriptstyle \Delta}{=}} \: }

\newfont{\boldlarge}{msbm10 scaled 1100}

\newcommand{\comment}[1]{}

\newlength{\tmpbigbar}
